\documentstyle[twocolumn,epsf,epsbox]{jpsj}
\def\bmit#1{\mbox{\boldmath$ #1 $}}
\def\runtitle{Complementarity relation of irreversible process}
\def\runauthor{Ken {\sc Sekimoto} and Shin-ichi {\sc Sasa}}

\title {Complementarity relation 
for irreversible process \\
derived from stochastic energetics}

\author {Ken {\sc Sekimoto}$^1$\footnote{e-mail:
sekimoto@yukawa.kyoto-u.ac.jp}
and 
Shin-ichi {\sc Sasa}$^2$\footnote{e-mail:
sasa@jiro.c.u-tokyo.ac.jp}}
\inst {$^1$Yukawa Institute for Theoretical Physics, Kyoto University, 
Kyoto 606-01, Japan and\\
 $^2$Department of Pure and Applied Science, University of Tokyo, 
Komaba, Meguro-ku 3-8-1 Tokyo 153, Japan}

\recdate {\today}

\abst {When the process of a system in contact with a heat bath
is described by classical
Langevin equation, the method of stochastic energetics 
[K. Sekimoto, {\it J. Phys. Soc. Jpn.} {\bf 66} (1997) 1234]
enables to derive the form of Helmholtz free energy and the
dissipation function of the system.
We prove that the irreversible heat $Q_{\rm irr}$ and the time lapse
${\Delta t}$ of an isothermal process obey the complementarity
relation, $ Q_{\rm irr} \, {\Delta t}
\geq k_{\rm B}T \, {\cal S}_{\rm min}$, where
$ {\cal S}_{\rm min}$ depends on the initial and the final 
values of the control parameters, but it does not depend on 
the pathway between these values.}

\kword {irreversible thermodynamics, complementarity, energetics, 
Langevin}

\begin{document}
\sloppy
\maketitle
\sloppy
The relations of thermodynamics tell that, when we change
quasi-statically
the control parameters of a system in contact with a heat bath,
the work $W$ needed for the change is equal to the change of
Helmholtz free energy, $\Delta F$, which consists of 
the reversible heat released to the heat bath, $Q_{\rm rev}$,
and the change of internal energy, $\Delta E$, as 
$Q_{\rm rev}+ \Delta E=\Delta F$.
When the change of the control parameter is not 
quasi-static, the needed work is more than the reversible one,
i.e., $W-\Delta F \geq 0$, and the released heat $Q$, which
obeys the energy conservation law $Q+\Delta E=W$,
is larger than $Q_{\rm rev}$ by
the amount so-called irreversible heat $Q_{\rm irr}
\equiv Q- Q_{\rm rev}=W-\Delta F$.

In order to assess the released heat $Q$ under a given protocol of
control parameters, we need both a dynamical model of a system and
a proper kinematical interpretation of
the heat release from the system.
One of the present authors~\cite{KS97} have introduced a method to
obtain $Q$ in the systems whose dynamics is described by Langevin equations.
We will show below that, if applied to slow change of
control parameters, this method, which we shall provisionally call
stochastic energetics, enables to formulate
reversible and irreversible thermodynamics of  processes
of the system in contact with a heat bath.

First we recapitulate the main idea of stochastic energetics
in the case of a single heat bath.
Let ${\mib x} = \{x_1, \ldots, x_n\}$ represent the state of the
fluctuating system and let ${\mib a}=\{ a_1, \ldots , a_r\}$
be the parameters which control the system through the potential 
$U({\mib x};{\mib a})$. The Langevin equation is assumed as follows,
\begin{equation}
{\bf 0}= -\Gamma \cdot \frac{d{\mib x}}{dt} + \bmit{\xi}(t) -
\frac{\partial U}{\partial {\mib x}}({\mib x};{\mib a}),
\label{langevin}
\end{equation}
where $\Gamma$ is a positive definite and symmetric friction matrix,
which we assume to be constant, and $\bmit{\xi}(t)$ is white and 
Gaussian random forces characterized by $\langle
\bmit{\xi}(t)\rangle
=0$ and $\langle \bmit{\xi}(t) 
\mbox{\raisebox{1.0ex}{\rm t}} \bmit{\xi}(t')\rangle=$ 
$2 \Gamma k_{\rm B}T \delta (t-t')$.~\cite{remark1}
As far as we can regard (\ref{langevin}) as a mechanical balance
equation, the balance equation for energy is obtained by making
the scalar products of each term with 
$d {\mib x}(t)\equiv {\mib x}(t+\frac{d t}{2}) -$
${\mib x}(t-\frac{d t}{2})$
along the {\it realized} trajectory, 
\begin{eqnarray}
&\hspace{-1cm}-( -\Gamma \cdot \frac{d{\mib x}}{dt}(t)+\bmit{\xi}(t)) \cdot d{\mib x}(t) 
\qquad \qquad
\nonumber \\
&\qquad =
\frac{\partial U}{\partial {\mib a}}({\mib x}(t);{\mib a}(t))
\cdot d{\mib a}(t) -dU({\mib x}(t);{\mib a}(t)),
\label{firstlaw}
\end{eqnarray}
where we have used  the identity
$dU= \frac{\partial U}{\partial {\mib x}}\cdot  d{\mib x}+$
$\frac{\partial U}{\partial {\mib a}}\cdot d{\mib a}$, which is valid 
for stochastic variables as far as the multiplication is defined in the 
Stratonovich sense.~\cite{Ga}

The left hand side is the heat released to the heat bath, $dQ$, since
$-( -\Gamma \frac{{\mib x}}{dt}(t)+\bmit{\xi}(t))$ is the reaction force
which the system exerts onto the heat bath. 
The energy conservation law implies that
$dQ +dU$ should be equal to 
the work $d{\cal W}$ done by the external agent to the
system, hence
\begin{equation}
d{\cal W} = \frac{\partial U}{\partial {\mib a}}({\mib x}(t);{\mib a}(t))
\cdot d{\mib a}(t).
\label{work}
\end{equation}
Suppose that the control parameters ${\mib a}$ is changed (by
some external agent) from ${\mib a}_{\rm i}$ at $t=0$ to 
${\mib a}_{\rm f}$ at $t={\Delta t}$.
The total work  $\cal W$ done to the system in the course of
 a particular process 
${\mib x}(t)$ ($0\leq t\leq {\Delta t}$) is then,
\begin{equation}
{\cal W}=\int^{\Delta t}_0 dt
\frac{\partial U}{\partial {\mib a}}
({\mib x}(t);{\mib a}(t))\cdot \frac{d {\mib a}(t)}{dt}.
\end{equation}
That the work $\cal W$ is defined {\it before} we take ensemble average
is both conceptual and practical advantage of stochastic energetics 
as compared with the master equation approaches.~\cite{Bergmann,Spohn}
The ensemble average of $\cal W$ over 
possible realization of $\{ \bmit{\xi}(t)\}_{0\leq t\leq {\Delta t}}\,$ 
is expressed as
\begin{eqnarray}
&   \hspace{-4cm}W\equiv \langle {\cal W}\rangle \qquad \qquad \nonumber \\ 
&\quad = \int^{\Delta t}_0 dt\,
\left[ \int d{\mib x} P({\mib x},t)\frac{\partial U}{\partial {\mib a}}
({\mib x};{\mib a}(t))\right] \cdot \frac{d {\mib a}(t)}{dt},
\label{avwork}
\end{eqnarray}
where $P$ is the probability distribution function of ${\mib x}$ which
obeys the following Fokker-Planck equation,~\cite{Ga}
\begin{eqnarray}
&\hspace{-3cm} \frac{\partial P}{\partial t}({\mib x},t)
=-{\cal L}_{\rm FP}({\mib a}(t)) P({\mib x},t) 
\nonumber \\
&
= \frac{\partial}{\partial {\mib x}}\cdot {\Gamma}^{-1} \cdot
\: \mbox{\raisebox{1.7ex}{\rm t}}\! \! \left(
 \frac{\partial U}{\partial {\mib x}}({\mib x};{\mib a}(t)) 
+ k_{\rm B}T \frac{\partial}{\partial {\mib x}}\right) P({\mib x},t). 
\label{FPeq}
\end{eqnarray}

\null

Below we describe the main results, whose derivation will be
given in the last part of the text.
Let $\hat{\mib a}(s)$ be a given protocol of the control parameter
${\mib a}$ with a unit time lapse (${\Delta t}=1$) satisfying  
$\hat{\mib a}(0)={{\mib a}}_{\rm i}$ and $\hat{\mib a}(1)={{\mib a}}_{\rm f}$.
We assume, for convenience, the unit of time to be the characteristic
time of equilibration of $ P({\mib x},t)$ when ${\mib a}$ is fixed
at a typical value of the protocol. 
Then for the stretched or slowed-down protocol 
${\mib a}(t)=$ $\hat{\mib a}(\frac{t}{{\Delta t}})$ with a large time lapse
${\Delta t}$, the following asymptotic formula for $W$ holds,
\begin{eqnarray}
W&=& \Delta F 
\nonumber \\
&&+ \frac{k_{\rm B}T}{{\Delta t}}
\int^1_0 ds 
\: \mbox{\raisebox{1.7ex}{\rm t}}\! \! \left(
  \frac{d \hat{\mib a}(s)}{ds} 
\right)
\cdot \Lambda(\hat{\mib a}(s))\cdot    \frac{d \hat{\mib a}(s)}{ds}
\nonumber \\
&& +{\cal O}({{(\Delta t)}^{-2}}),
\label{asymptot}
\end{eqnarray}
where $\Delta F$ is the difference of the Helmholtz free energy,
$\Delta F \equiv$ $F({\mib a}_{\rm f})-$ $F({\mib a}_{\rm i})$ with
\begin{equation}
F({\mib a})\equiv  -k_{\rm B}T \: \log
\left\{ \int d{\mib x}\: \exp\left[ 
-\frac{U({\mib x};{\mib a})}{k_{\rm B}T}\right]
\right\} + \mbox{const.}
\label{freeenergy}
\end{equation}
and $\Lambda({\mib a})$ is a positive definite $n\times n$ matrix
defined by
\begin{eqnarray}
&\hspace{-4.5cm} \Lambda({\mib a})=-
\int \!d{\mib x}\! \int \! d{\mib x}'
\nonumber \\
&\frac{\partial P_{\rm eq}}{\partial {\mib a}}({\mib x};{\mib a})
{g}({\mib x},{\mib x}';{\mib a})
\: \mbox{\raisebox{1.7ex}{\rm t}}\! \! \left(
\frac{\partial P_{\rm eq}}{\partial {\mib a}}
({\mib x}';{\mib a})\right),
\label{kernel}
\end{eqnarray}
where $P_{\rm eq}({\mib x};{\mib a})\equiv$ 
$e^{-U({\mib x};{\mib a})/{k_{\rm B}T}}/
\int d\bar{\mib x} e^{-U(\bar{\mib x};{\mib a})/{k_{\rm B}T}}$
is the equilibrium distribution under a given parameter value,
${\mib a}$,
and the kernel ${g}({\mib x},{\mib x}';{\mib a})$ 
is defined as the solution of the following equation 
\begin{equation}
-{\cal L}_{\rm FP}({\mib a})
[{P_{\rm eq}({\mib x};{\mib a})} g({\mib x},{\mib x}';{\mib a})]=
\delta({\mib x}-{\mib x}').
\label{green}
\end{equation}

The result (\ref{asymptot}) tells firstly that,
in the limit of slow and smooth change of external parameters
(${\Delta t} \to \infty$), the stochastic energetics reproduces 
correctly the thermodynamic relation of quasi-static isothermal 
processes, $W=\Delta F$.
Secondly, for large but  finite time lapse ${\Delta t}$, the irreversible heat
$Q_{\rm irr} = W-\Delta F$ behaves asymptotically as $\sim
\frac{1}{{\Delta t}}$, with the proportionality constant depending on
the scaled protocol $\hat{\mib a}(\frac{t}{{\Delta t}})$.
The integral on the right hand side of  (\ref{asymptot}) is
analogous to an classical action of free particle with 
a `mass' $2\Lambda$ being dependent on the `position' ${\mib a}$.
The minimum value of this integral, which we 
denote by ${\cal S}_{\rm min}({\mib a}_{\rm i},{\mib a}_{\rm f})$,
 should be realized by a certain `classical' path, or the optimal (scaled) 
protocol $\hat{\mib a}^*(s)$ ($0\leq s
\leq 1$). We then come to the following complementarity relation
which is correct asymptotically for ${\Delta t} \to \infty$,
\begin{equation}
 Q_{\rm irr}\, {\Delta t} \geq k_{\rm B}T\, 
{\cal S}_{\rm min}({\mib a}_{\rm i},{\mib a}_{\rm f})
\label{compl}
\end{equation}
The above relation implies a sort of uncertainty relation 
that the estimation of the Helmholtz free energy function by the 
measurement of mechanical work $W$ inevitably includes a deviation 
$Q_{\rm irr}$ whose lower bound is controlled by the inverse of 
the measuring time $\Delta t$, at least in moderately slow measurements.

\underline{\it Remark 1.}$\:$
 The dissipation function $\Phi$ of linear irreversible 
thermodynamics is usually defined as $\frac{1}{T} \frac{dQ_{\rm irr}}{dt} =$
$2  \Phi( \frac{d {\mib a}(t)}{dt}).$
In our method $\Phi$ is obtained 
if we neglect the ${\cal O}({(\Delta t)}^{-2})$ 
terms in (\ref{asymptot}) and rewrite the integral there using 
${\mib a}(t)$ instead of $\hat{\mib a}(s)$;
\begin{equation}
\Phi( \frac{d {\mib a}(t)}{dt})
=  \frac{k_{\rm B}}{2}\,
\mbox{\raisebox{1.7ex}{\rm t}}\! \! \left(\frac{d {\mib a}(t)}{dt}\right)
\cdot \Lambda({\mib a}(t))\cdot \frac{d {\mib a}(t)}{dt}.
\label{dissipationfunction}
\end{equation}

\underline{\it Remark 2.}$\:$ 
In terms of the spectral representation of the Fokker-Planck
operator, 
\begin{equation}
{\cal L}_{\rm FP}({\mib a})=
\sum^{\infty}_{m=0} 
|m;{\mib a}\rangle {\lambda_m({\mib a})}\langle m; {\mib a}|
\end{equation}
with $0=\lambda_0({\mib a}) < \lambda_1({\mib a})\leq $ 
$\lambda_2({\mib a})\leq \ldots$ and
$\langle m;{\mib a}|n;{\mib a}\rangle = \delta_{mn}$,
the kernel $\Lambda({\mib a})$ is formally expressed as
\begin{eqnarray}
&\hspace{-3.5cm}
\Lambda({\mib a})=\frac{1}{{(k_{\rm B}T)}^2}\times \qquad \qquad
\nonumber \\
& \sum^{\infty}_{m=1} 
\langle 0;{\mib a}|
\: \mbox{\raisebox{1.7ex}{\rm t}}\! \! \left(
\frac{\partial U}{\partial {\mib a}}
\right)
|m;{\mib a}\rangle \frac{1}{\lambda_m({\mib a})}\langle m; {\mib a}|
\frac{\partial U}{\partial {\mib a}}
|0;{\mib a}\rangle.
\label{kernel3}
\end{eqnarray}
Here ${\mib a}$ appears merely as a parameter.

\underline{\it Remark 3.}$\:$ 
The function $g({\mib x},{\mib x}';{\mib a})$ 
is the Green's function of a Hermitian operator; 
\begin{equation}
k_{\rm B}T \frac{\partial}{\partial {\mib x}}\cdot
{\Gamma}^{-1}
 P_{\rm eq}({\mib x};{\mib a})\cdot
 \mbox{\raisebox{1.7ex}{\rm t}}\! \! \left(
\frac{\partial}{\partial {\mib x}} 
\right) g({\mib x},{\mib x}';{\mib a})=\delta ({\mib x}-{\mib x}').
\end{equation}
Especially when ${\mib x}$ is a single variable, $x$, and
$\Gamma$, a constant $\gamma$, the explicit form of $g(x,x';{\mib a})$ is 
\begin{equation}
g(x,x';{\mib a})=\frac{\gamma}{2 k_{\rm B}T} \mbox{sgn}(x-x')
\int^x_{x'} \frac{dz}{P_{\rm eq}(z;{\mib a})}.
\end{equation}

\underline{\it Remark 4.}$\:$ 
If the control parameters ${\mib a}$ are constrained
to change along a single trajectory, say $\tilde{\mib a}(\theta)$,
where $0\leq \theta \leq 1$ is the contour parameter with 
 $\tilde{\mib a}(0)=$ ${\mib a}_{\rm i}$ and 
$\tilde{\mib a}(1)=$ ${\mib a}_{\rm f}$, then $Q_{\rm irr}$ 
depends yet on how $\theta$ depends on the scaled time $s$, 
as well as on the time lapse ${\Delta t}$. The integral in
(\ref{asymptot}) then becomes 
$\int^1_0 \Lambda(\theta(s)) {\left| \frac{d\theta}{ds}(s)\right|}^2 ds,$
where 
$\Lambda(\theta)\equiv  \: \mbox{\raisebox{1.0ex}{\rm t}}\! \! 
\left(  \frac{d \tilde{\mib a}(\theta)}{d\theta} \right)
 \cdot \Lambda(\tilde{\mib a}(\theta))\cdot   \frac{d
\tilde{\mib a}(\theta)}{d\theta}$.
If we note that this integral becomes the simple action 
$\int^1_0 {\left| \frac{du}{ds}(s)\right|}^2\, ds$ 
by the transformation $du(s) \equiv {\Lambda(\theta(s))}^{1/2} d\theta(s) $,
the minimum of the integral is explicitly given as
${\left| \int^1_0 {\Lambda(\theta)}^{1/2} d\theta \right| }^2$.
Although not apparent, this result is invariant under the relabeling 
of the contour parameter.

\null

As an illustration, let us apply our formulae to the case of 
$U(x;a)=\frac{a}{2} x^2$ with the single variable $x$ and the
single control parameter $a$, with $\Gamma$ a constant $\gamma$.
In  ${\cal O}({\Delta t}^0)$ order we recover the
correct (configurational) free energy
\begin{equation}
\Delta F= \frac{k_{\rm B}T}{2}\log\left( \frac{a_{\rm f}}{a_{\rm i}}\right).
\end{equation}
In this case $F$ is purely of entropic origin since, from 
equipartition theorem, the internal energy is independent of the
`width' of the potential, $\sim 1/\sqrt{a}$.

In ${\cal O}({(\Delta t)}^{-1})$ order, we obtain the irreversible heat
\begin{equation}
Q_{\rm irr}=
\frac{\gamma\, k_{\rm B}T}{4\Delta t}
\int^1_0 ds \frac{1}{{\hat{a}(s)}^3}
{\left| \frac{d \hat{a}(s)}{ds}\right|}^2
+{\cal O}({(\Delta t)}^{-2}).
\end{equation}
The integral can be minimized by the following scaled protocol;
\begin{equation}
\hat{a}(s)|_{\rm optimum}={\left| 
\frac{s}{\sqrt{a_{\rm f}}}+\frac{1-s}{\sqrt{a_{\rm i}}}
\right|}^{-2},
\end{equation}
and the asymptotic complementarity relation reads as follows,
\begin{equation}
Q_{\rm irr} \Delta t \geq \frac{\gamma \, k_{\rm B}T}{4}
{\left| \frac{1}{\sqrt{a_{\rm i}}}- \frac{1}{\sqrt{a_{\rm f}}} 
\right| }^2.
\end{equation}

\null

\underline{\it Proof of (\ref{asymptot})}$\:$
If we introduce the scaled time $s=t/\Delta t$ and the probability 
distribution with this argument, $\hat{P}({\mib x},s;\Delta t)\equiv 
P({\mib x},s\Delta t)$, the equations (\ref{avwork}) and(\ref{FPeq})
becomes, respectively,
\begin{equation}
W=\int^1_0 ds  \frac{d \hat{\mib a}(s)}{ds}\cdot \int d{\mib x} 
\frac{\partial U}{\partial {\mib a}}({\mib x};\hat{\mib a}(s))
 \hat{P}({\mib x},s;\Delta t)
\label{avwork2}
\end{equation}
\begin{equation}
\frac{1}{\Delta t}\frac{\partial \hat{P}}{\partial s}({\mib x},s;\Delta t)
= -{\cal L}_{\rm FP}(\hat{\mib a}(s)) \hat{P}({\mib x},s;\Delta t).
\label{FPeq2}
\end{equation}
For large $\Delta t$, we may solve (\ref{FPeq2}) perturbatively 
in the form of
\begin{equation}
\hat{P}({\mib x},s;\Delta t)
=\hat{P}^{(0)}({\mib x},s) +
\frac{1}{\Delta t} \hat{P}^{(1)}({\mib x},s)+
\cdots.
\end{equation}
In the lowest order $\hat{P}^{(0)}$ obeys
 ${\cal L}_{\rm FP}(\hat{\mib a}(s))\hat{P}^{(0)}({\mib x},s)=0$, 
and the normalization condition
$\int d{\mib x} \hat{P}^{(0)} ({\mib x},s)=1$. 
Then, $\hat{P}^{(0)} ({\mib x},s)$ is the equilibrium
distribution for a given parameter  $\hat{\mib a}(s)$, i.e.,
\begin{equation}
\hat{P}^{(0)} ({\mib x},s)=P_{\rm eq}({\mib x};\hat{\mib a}(s)).
\end{equation} 
From (\ref{avwork}) and the identity 
$\int d{\mib x}\frac{\partial U}{\partial {\mib a}}({\mib x};{\mib a}) 
P_{\rm eq}({\mib x};{\mib a})=$
$\frac{\partial F}{\partial {\mib a}}({\mib a}),$
the ${\cal O}({\Delta t}^0)$ term of $W$ becomes $F({\mib a}_{\rm
f})-$ $F({\mib a}_{\rm i})$.

In the next order, (\ref{FPeq2}) becomes,
$-{\cal L}_{\rm FP}(\hat{\mib a}(s))\hat{P}^{(1)}({\mib x},s)=$
$\frac{\partial}{\partial s}P_{\rm eq}({\mib x},\hat{\mib a}(s)).$
Using (\ref{green}), the solution is given as
\begin{eqnarray}
&\!\!\! 
\hat{P}^{(1)}({\mib x},s)=P_{\rm eq}({\mib x};\hat{\mib a}(s))
\nonumber \\
&\times\left[
\int d{\mib x}' g({\mib x},{\mib x}';\hat{\mib a}(s))
\frac{\partial P_{\rm eq}}{\partial s}({\mib x}';\hat{\mib a}(s)) +\chi
\right],
\end{eqnarray}
where $\chi$ is to be determined from the normalization condition, 
$\int d{\mib x} \hat{P}^{(1)}({\mib x},s)=0$.
After performing this, and noting the relation 
$\frac{\partial P_{\rm eq}}{\partial s}({\mib x}';\hat{\mib a}(s))=$
$ \mbox{\raisebox{1.7ex}{\rm t}}\! \! \left(
\frac{\partial P_{\rm eq}}{\partial {\mib a}}({\mib x}';\hat{\mib a}(s))
\right)  \cdot \frac{d \hat{\mib a}(s)}{ds},$
the kernel $\Lambda(\hat{\mib a}(s))$ in (\ref{asymptot}) becomes as follows,
\begin{eqnarray}
\Lambda({\mib a})&=
\frac{1}{k_{\rm B}T}\int d{\mib x} \int d{\mib x}'
\: \mbox{\raisebox{1.7ex}{\rm t}}\! \! \left(
\frac{\partial U}{\partial {\mib a}}({\mib x};{\mib a}) 
\right)
P_{\rm eq}({\mib x};{\mib a})
\nonumber \\
& \times
[{\cal P}^\perp_{{\mib x}}({\mib a}) 
g({\mib x},{\mib x}';{\mib a})]
\frac{\partial P_{\rm eq}}{\partial {\mib a}}({\mib x}';{\mib a}).
\label{kernel2}
\end{eqnarray}
where we have introduced an operator 
${\cal P}^\perp_{\mib x}({\mib a})$ 
such that 
\begin{equation}
{\cal P}^\perp_{\mib x}({\mib a}) \psi({\mib x})\equiv
 \int d\bar{\mib x}
[\delta(\bar{\mib x}-{\mib x})-P_{\rm eq}(\bar{\mib x};{\mib a})]
{\psi}(\bar{\mib x})
\label{transverse}
\end{equation}
for an arbitrary function of $\psi({\mib x})$. 
Next the factors $ \mbox{\raisebox{1.7ex}{\rm t}}\! \! \left(
\frac{\partial U}{\partial {\mib a}}({\mib x};{\mib a}) 
\right)$ $ P_{\rm eq}({\mib x};{\mib a})$ in (\ref{kernel2}) can be rewritten as
\begin{eqnarray}
& \mbox{\raisebox{1.7ex}{\rm t}}\! \! \left(
\frac{\partial U}{\partial {\mib a}}({\mib x};{\mib a}) 
\right) P_{\rm eq}({\mib x};{\mib a})=
-k_{\rm B}T 
\frac{\partial P_{\rm eq}}{\partial {\mib a}}({\mib x};{\mib a})
\nonumber \\
&\qquad +P_{\rm eq}({\mib x};{\mib a})
\int d\bar{\mib x} 
\mbox{\raisebox{1.7ex}{\rm t}}\! \! \left(
\frac{\partial U}{\partial {\mib a}}(\bar{\mib x};{\mib a}) 
\right) P_{\rm eq}(\bar{\mib x};{\mib a}).
\end{eqnarray}
Here the second term on the right hand side 
($\propto~P_{\rm eq}({\mib x};{\mib a})$) does not contribute to 
the integral of (\ref{kernel2}) since the following identity
holds for an arbitrary function,  $\psi({\mib x})$,
\begin{equation}
\int d{\mib x} 
{P_{\rm eq}}({\mib x};{\mib a})
[{\cal P}^\perp_{{\mib x}}({\mib a})  \psi({\mib x}) ]
=0.
\end{equation}
Now the integral in $\Lambda({\mib a})$ is of the form of
$\int d{\mib x}
\frac{\partial P_{\rm eq}}{\partial {\mib a}}({\mib x};{\mib a})
 \psi({\mib x}),$
and since the following identity
\begin{eqnarray}
&\hspace{-3cm}
\int d{\mib x} 
\frac{\partial P_{\rm eq}}{\partial {\mib a}}({\mib x};{\mib a})
[{\cal P}^\perp_{\mib x}({\mib a})  \psi({\mib x}) ]
\nonumber \\
&=
\int d{\mib x}
\frac{\partial P_{\rm eq}}{\partial {\mib a}}({\mib x};{\mib a})
 \psi({\mib x}) 
\end{eqnarray}
holds for an arbitrary function $\psi({\mib x})$,
we come to the expression (\ref{kernel}).

\null

In this letter we have considered the irreversible process near
equilibrium states. 
Although it may not have been explicitly stated, the complementarity 
relation itself is derivable from phenomenological dissipation function. 
The method of stochastic energetics not only gives a concrete form of 
the bound of the uncertainty between $Q_{\rm irr}$ and $\Delta t$,
but it is also applicable to not slow processes such as thermal 
ratchets~\cite{KS97} or far from equilibrium non-steady states.~\cite{KS8} 
How the irreversible heat does or does not depend on the models 
under the non-slow processes is a challenging problem in 
future.~\cite{Jarzynski}

\acknowledgements
The authors gratefully acknowledges M. Tokunaga, K. Sato, A. Tanaka, T. Ono, 
T. Sasada, H. Hasegawa, and Y. Oono for valuable comments.
This work was partly supported by the grants from the Ministry of
Education, Science, Sports and Culture of Japan, Nos. 09279222,
 09874079 (KS) and 09740305 (SS), from Asahi Glass fundation (KS) 
and from National Science Foundation, No. NSF-DMR-93-14938 (SS).

\end{document}